\documentclass[a4paper]{jpconf}
\usepackage{graphicx}
\usepackage{color}
\usepackage{amssymb}
\usepackage{amsmath}
\usepackage{multirow}
\usepackage{amscd}
\usepackage{verbatim}

\definecolor{cream}{rgb}{.97, .95, .88}
\definecolor{darkcream}{rgb}{1., .88, .5}
\definecolor{lightpink}{rgb}{0.98, 0.88, 0.87}
\definecolor{lightwhite}{rgb}{1., 0.98, 0.95}
\definecolor{lightsalmon}{rgb}{1., 0.95, 0.90}
\definecolor{lightviolet}{rgb}{0.9, 0.8, 0.9}
\definecolor{lightgray}{rgb}{.96, .96, .96}  
\definecolor{lgray}{rgb}{.75, .75, .75}
\definecolor{LemonChiffon}{rgb}{0.95, 1., 0.7}
\definecolor{lightolivegreen}{rgb}{0.84, 0.89, 0.25}
\definecolor{lightgreen}{rgb}{.664, 1., .52}
\definecolor{llgreen}{rgb}{.900, .983, .960}
\definecolor{tristle}{rgb}{0.87, 0.67, 0.77} %{0.792, 0.609, 0.698}
\definecolor{pink}{rgb}{0.95, 0.45, 0.75}
\definecolor{magenta}{rgb}{1., 0, 1.}
\definecolor{violet}{rgb}{0.9, 0.20, 0.85}
\definecolor{darkolivegreen}{rgb}{0.55, 0.65, 0.35}
\definecolor{maroon}{rgb}{0.7, 0.26, 0.56}
\definecolor{lightmaroon}{rgb}{0.85, 0.38, 0.58}
\definecolor{darkmaroon}{rgb}{0.604, 0.169, 0.451}
\definecolor{ddarkmaroon}{rgb}{0.2, 0.03125, 0.150}
\definecolor{mediumorchid}{rgb}{0.8, 0.33, 0.83}
\definecolor{mediumorchidd}{rgb}{1., 0.33, 0.63}
\definecolor{darkgreen}{rgb}{0.1, 0.6, 0.13}
\definecolor{lightyellow}{rgb}{1., 1., 0.82}
\definecolor{turquoise}{rgb}{0.042, 0.586, 0.512}
\definecolor{turquoisel}{rgb}{0.66, 0.94, 0.83}
\definecolor{darkturquoise}{rgb}{0.21, 0.55, 0.50}
\definecolor{coral}{rgb}{1., 0.6, 0.21}
\definecolor{lightorange}{rgb}{1., 0.88, 0.75}
\definecolor{orangered}{rgb}{1., 0.5, 0.}
\definecolor{orange}{rgb}{1., 0.65, 0.1}
\definecolor{orangel}{rgb}{1., .85, .3}
\definecolor{darkorange}{rgb}{0.875, 0.4, 0.204}
\definecolor{ddarkorange}{rgb}{.675, .218, .05}
\definecolor{bluesky}{rgb}{0.48, 0.53, 1.}
\definecolor{gold}{rgb}{1., 0.85, 0.25}
\definecolor{goldd}{rgb}{0.95, 0.75, 0.05}
\definecolor{darkviolet}{rgb}{0.54, 0.04, 0.84}
\definecolor{ddarkviolet}{rgb}{.382, .063, .657}
\definecolor{lightblue}{rgb}{0.30, 0.86, 0.89}
\definecolor{LightBlue}{rgb}{0.68, 0.85, 0.9}
\definecolor{lblue}{rgb}{0.78, 0.90, 0.95}
\definecolor{darkblue}{rgb}{.105, .308, .707}
\definecolor{lightmaroon}{rgb}{0.85, 0.38, 0.58}
\definecolor{darkmaroon}{rgb}{0.604, 0.169, 0.451}
\definecolor{darkpink}{rgb}{0.879, 0.020, 0.766}
\definecolor{ddarkpink}{rgb}{0.738, 0.195, 0.406}
\definecolor{grey}{rgb}{0.717, 0.717, 0.717}
\definecolor{lightgrey}{rgb}{0.800, 0.800, 0.800}
\definecolor{brown}{rgb}{0.740, 0.323, 0.182}
\definecolor{redbrown}{rgb}{.575, .158, .05}
\definecolor{darkbrown}{rgb}{0.34, 0.25, 0.05}
\definecolor{orangebrown}{rgb}{0.433, 0.262, 0.06}
\definecolor{pinkl}{rgb}{1., 0.788, 0.918}
\definecolor{salmon}{rgb}{1., 0.66, 0.5}
\definecolor{lightbrown}{rgb}{0.703, 0.508, 0.121}

\def\Journal#1#2#3#4{#3 {#1} {\bf #2} #4}

\def\etal{{\it et al.}}
\def\JPS{\em J. Phys. Chem. Solids}

\def\MPA{{\em Mod. Phys.} A}

\def\NAC{\em Nature Commu.}

\def\NAT{\em Nature}

\def\PHY{\em Physics}

\def\PRD{{\em Phys. Rev.} D}
\def\PRL{\em Phys. Rev. Lett.}
\def\PRV{\em Phys. Rev.}

\def\RMP{\em Rev. Mod. Phys.}

\def\be{\begin{equation}}
\def\ee{\end{equation}}
\def\bea{\begin{eqnarray}}
\def\eea{\end{eqnarray}}
\def\bes{\begin{equation*}}
\def\ees{\end{equation*}}
\def\beas{\begin{eqnarray*}}
\def\eeas{\end{eqnarray*}}
\begin{document}
\title{Symmetry as a foundational concept in Quantum Mechanics}
\author{Houri~Ziaeepour}
\address{Institut UTINAM-CNRS, Universit\'e de Franche Comt\'e, Besan\c{c}on, France}
\ead{houriziaeepour@gmail.com}

\begin{abstract}
Symmetries are widely used in modeling quantum systems but they do not contribute in postulates 
of quantum mechanics. Here we argue that logical, mathematical, and observational evidence 
require that symmetry should be considered as a fundamental concept in the construction of 
physical systems. Based on this idea, we propose a series of postulates for describing quantum 
systems, and establish their relation and correspondence with axioms of standard quantum 
mechanics. Through some examples we show that this reformulation helps better understand 
some of ambiguities of standard description. Nonetheless its application is not limited to 
explaining confusing concept and it may be a necessary step toward a consistent model of quantum 
cosmology and gravity.

\end{abstract}

\section{Introduction}
Looking at the history of science and technology of $20^{th}$ century and these early years of 
$21^{st}$ century, we find that most of them are directly or indirectly the fruit of understanding 
and application of quantum mechanics. From electronic components in smart phones which have become 
an indispensable gadget in human life, to electricity generated by nuclear or solar energy 
facilities, to understanding of photosynthesis phenomenon and maybe its artificial imitation 
in future, couldn't be discovered, understood, controlled, and applied without quantum mechanics. 

However, despite owing so much to this most fundamental and profound law of physics, and despite 
its success in harshest and most stringiest tests that present science and technology allow, it 
is yet considered by many, including some of participants of this conference\footnote{\it $7^{th}$ 
International Workshop DICE2014, Spacetime - Matter - Quantum Mechanics, Sep. 15-19, 2014, 
Castigilioncello, Tuscany, Italy.}, to be incomplete and only the state of knowledge rather than 
complete reality. For this reason since its formulation in 1930 by Dirac and von Neumann, many 
alternative interpretations of quantum phenomena are suggested. Some of these alternatives such as 
deterministic interpretations and local hidden variables are ruled 
out~\cite{bellexperiments,bellexperiments0} - although there are claims that not all loop holes are 
closed and deterministic probably testable extension of quantum mechanics 
exist~\cite{hidenvarglobal}. Furthermore, many of alternative interpretations do not present a 
discriminating test and in this regard are equivalent to quantum mechanics but less natural and 
more obscure. Example of these models are Bohmian quantum mechanics~\cite{bohemqm} and 
Everett {\it many worlds} interpretation~\cite{qmmanyworld}.

In this proceedings we review a new proposal~\cite{houriqmsymm} for reformulating axioms of the 
standard quantum mechanics - rather than a new interpretation - using the logical and physical 
concept of symmetry. We show that some of rules such as Born law for determination of expectation 
values can be obtained in place of being postulated and introduce by hand in the theory. Moreover, 
in this new description confusing concepts such as nonlocality and measurement problem can be 
better explained and understood. Indeed, in modeling quantum systems and phenomena, in particular 
in particle physics and condensed matter, symmetries are extensively used and play a central role 
in mathematical formulation and physical interpretation. But they do not explicitly appear in 
quantum foundation and postulates as suggested by Dirac and von Neumann. In mathematical logic, 
symmetry can be considered as an extension of fundamental logical concept of equality. It 
describes and quantifies degeneracies and maximum amount of information carried by a system in an 
abstract manner and independent of details.

In the following sections we first present axioms of quantum mechanics according to this new 
description. Then, through mathematical demonstration and case examples we show that the theory 
obtained from these axioms is equivalent to the standard formulation of quantum axioms. 
Nonetheless, confusing quantum observations acquire straightforward logical and mathematical 
interpretation directly related to symmetries.

\section{Quantum mechanics postulates in symmetry language}
\renewcommand{\theenumi}{\roman{enumi}}
\begin{enumerate}
\item A quantum system is defined by its symmetries. Its state is a vector belonging to a 
projective vector space -called state space - which represents symmetry groups of the system. 
Observables are associated to self-adjoint operators and a set of independent observables is 
isomorphic to a subspace of commuting elements of the space of self-adjoint operators acting on 
the state space. 
\label{poststate}
\medskip
\item The state space of a composite system is homomorphic to the direct product of state spaces 
of its components. For non-interacting (separable) components, this homomorphism becomes an 
isomorphism. \label{postcomposite}
\item Evolution of a system is unitary and ruled by conservation laws imposed by its symmetries 
and their representation by state space. \label{postunitary}
\medskip
\item Decomposition coefficients of a state to eigen vectors\footnote{To be more precise we 
should use the term {\it rays} in place of vectors because vectors differing by a constant are 
equivalent. Thus, we assume that states are normalized.} of an observable presents the 
symmetry/degeneracy of the system with respect to its environment according to that observable. 
Its measurement is by definition the operation of breaking this symmetry/degeneracy. The 
outcome of the measurement is the eigen value of the eigen state to which the symmetry is broken. 
This spontaneous\footnote{We explain in Sec. \ref{sec:symmbrdeco} why the breaking of degeneracy 
between states should be classified as spontaneous.} symmetry breaking reduces the state space 
(the representation) to subspace generated by other independent observables.\label{postmeasure}
\medskip
\item A probability independent of measurement details is associated to eigen values of 
an observable as the outcome of a measurement. It presents the amount of symmetry/degeneracy of 
the state before its breaking by the measurement process. \label{postsymmbr}
\end{enumerate}

\section{Weirdnesses and unsettled issues of quantum mechanics}
Before describing how the new description of quantum mechanics helps understand microscopic world 
and quantum phenomena, we briefly review observations and concepts which seem confusing and weird:

\paragraph*{Complete but random}
The concept of randomness in physics existed before the discovery of quantum mechanics. But it 
was considered that it is an apparent rather than inherent and basic rule induced by inability 
of observing and adding to (classical) mechanical formulation of all degrees of freedom. 
Therefore, the fact that quantum mechanics only predicts a probability for outcomes of 
measurements and observations has been considered to be due to incompleteness of our information 
about physical world and its rules rather than fundamental. Consequently, {\it Hidden variables} 
were introduced to {\it complete} the information, and remove the randomness. However, 
observations, notably validation~\cite{bellexperiments} of Bell inequalities~\cite{bellineq}, 
have ruled out the presence of a local hidden physics. There yet claims of nonlocal hidden 
physics~\cite{hidenvarglobal,qmmanyworld} but there is no proposition how their existence can be 
proved.
\paragraph*{Nonlocality}
Classical world is local and and causal. In any process there is a well distinction between cause 
and effect. Even in Einstein relativity where order of events can be different for spacelike 
observers, causality is preserved for timelike observers who can exchange a signal with the 
limited and constant speed of light in vacuum. In quantum mechanics there are cases where cause 
and effect cannot be disentangled. This means that physical systems can be entangled and 
{\it communicate} with each other seemingly instantaneously. Einstein called this phenomenon 
{\it the spooky action at a distance}.
\paragraph*{Contextual}
In classical mechanics a random process stays random for ever. A coin bouncing back by an 
elastic surface has an observable and definitive state at the time of touching the surface, i.e. 
its initial state before bouncing can be known. Nonetheless its state in the next touch-down is 
again random and observation or non-observation of its initial state does not change the 
probability of outcomes in the next observation/measurement. The same is not true in quantum 
mechanics in which once an observable is measured, repetition of the operation gives exactly the 
same value. In addition, in the case of vector quantities, in some situations it is not clear 
which component is observed. This contextuality of quantum mechanics raises a number of questions:
\begin{description}
\item {- } Is a quantum state a physical reality ? If yes, why does it seem to have a non-unitary 
evolution when it is measured ?
\item {- } Why do measured quantities behave classically ? 
\item {- } Does the wave function collapse? 
\end{description}
In the following sections we describe answer to these questions in the framework of the 
reformulation of axioms of quantum mechanics based on symmetries.

\section{State space and its properties}
Symmetries are our only tool for abstraction of the Universe, and their breaking is the only 
logical and physical mean for distinguishing between objects. This observation is abstracted in 
representation of symmetry groups by vector spaces. Therefore, in contrast to the standard 
description of quantum mechanics which introduces an abstract state (Hilbert) space without 
specifying its relation with the system, postulate (\ref{poststate}) explicitly associates it 
to the representation of symmetries of the physical system. This space presents, in a unique 
manner, all the properties of a system and what can be learned about them. Of course since the 
discovery of quantum mechanics vector spaces representing symmetries of systems were used as 
the abstract Hilbert space mentioned in the quantum mechanics axioms \`a la Dirac and 
von Neumann, but the axioms do not clarify and impose this choice. Consequently, it stays unclear 
whether another space - {\it a more complete space} - can exist which explains properties of the 
system in a deterministic rather than random manner. The reformulated axiom (\ref{poststate}) 
emphasizes on the fact that only symmetries are meaningful and clarifies the abstract and 
unspecified nature of the state space. We remind that this postulate does not assume that the 
vector space of symmetry representation is a Hilbert space. We will prove this property in section 
(\ref{sec:measure}).

Like any vector space the state space is generated by eigen vectors of its commuting subgroup, 
which according to axiom (\ref{poststate}) defines independent observables. Decomposition 
of an arbitrary state vector depends on the selected basis, but a change of basis only modifies the 
decomposition of states not their identity or nature as a distinguishable member of the maximal 
abelian subalgebra of the associated algebra of the symmetry group. Therefore, in this description 
observables have an identity independent of the chosen basis. This fact is important, e.g. for 
interpretation of double Stern-Gerlach experiment, see the next subsection. 

\subsection{Active manifestation of symmetries in quantum mechanics and contextuality}
In contrast to classical mechanics, in quantum systems symmetries have an operative role and induce 
degeneracies which reflect history and relation of a system with its environment. This solves 
preferred basis issue of quantum mechanics. A transformation by $U \in G$ where $G$ is the 
symmetry group leads to:
\bes
\hspace{-1cm}|\psi\rangle = \sum_\alpha a_\alpha |\alpha\rangle,~|\alpha\rangle 
\rightarrow |\alpha'\rangle = U |\alpha \rangle,~\Longrightarrow~|\psi\rangle \rightarrow 
|\psi'\rangle = \sum_{\alpha'} a_\alpha |\alpha'\rangle
\ees
Therefore the maximal abelian subspace is projected to a new subspace. In general 
$|\alpha'\rangle$ are no longer eigen vectors of the new abelian subspace and coefficients 
$a_{\alpha'}$ do not have the same value and physical interpretation as $a_{\alpha}$. For instance, 
in a double Stern-Gerlach experiment electrons are polarized in the first detector and only 
one of polarizations, e.g. $Z^+$ is kept. We identify $\{|\alpha\rangle\} \equiv \{|Z^+\rangle, 
|Z^-\rangle\},~a_{Z^+} = 1,~a_{Z^-} = 0$. The passage of 
electrons through the second detector with a field in $X$ direction is equivalent to application 
of a symmetry transformation which projects the abelian subspace to $X$ direction. According to 
postulate (\ref{poststate}) and in agreement with observations, after passage to the second 
detector, observables belong to the new abelian subspace, and both polarizations in $X$ direction 
become possible. Therefore, the origin and naturalness of contextuality is the fact that physical 
systems actively present symmetries. 

Noncommuting position and momentum which both represent translation symmetry are considered to be 
example of preferred basis because it seems that collapsed/projected wave function of macroscopic 
objects prefers position basis. This misinterpretation has even led to construction of quantum 
gravity models based on position as preferred observable~\cite{causalset}. This apparent 
preference is evidently a misinterpretation and caused by the fact that most objects are 
non-relativistic with respect to us, and $\langle \hat{X}(t) - \hat{X}(t_0) \rangle \approx 0$. 
Thus, they look to be in a position eigen state. On the contrary, photons look like to prefer 
momentum basis. Therefore, preference of a basis is just a misconception. 

\section{State and symmetries in quantum and classical systems} \label{sec:state}
Degeneracies of a system determine the representation realized by its state space and induce a 
symmetry that we call {\it state symmetry}. For instance, in a hydrogen atom the Lagrangian has 
$O(3) \cong SU(2)$ symmetry, but electron state space is generated by $(r, \phi, \theta) 
\in O(3) \mathbb{T} \times \mathbb{T} \times \mathbb{T} \cong U(1) \times U(1) \times U(1) \subset 
\mathbb R^3 \supset O(3)$. Once an observable is {\it measured}, according to axiom 
(\ref {postmeasure}) the state symmetry or degeneracy between eigen vectors breaks, and thereby 
the representation is reduced to trivial. Therefore, state symmetry is related to systems history, 
initial condition, and preparation, presumably it is ephemeral and may be total i.e. all states 
have similar probability, or partial i.e. some are preferred to others. In the former case 
coefficients of decomposition of the state to eigen vectors are equal, otherwise the symmetry is 
partial, meaning that the system's history or environment somehow discriminates between eigen 
states but cannot single out one of them. The symmetry of states is by nature associated to 
system-environment symmetries and is inseparable from it. Therefore, breaking of degeneracies 
is comparable and analogue to spontaneous symmetry breaking in which the symmetry is broken in 
states but not (or softly) in the Lagrangian. Moreover, if state space is decomposed to orthogonal 
subspaces, breaking of symmetry in one block does not affect others. Notably, if a symmetry 
orthogonal to spacetime is broken, the effect will be global and in contrast to classical physics, 
spacetime is not treated in a special manner. This explains the origin of nonlocality in quantum 
mechanics and why there is no classical analogue for it.

Only in vector spaces defined over complex numbers $\mathbb{C}$, operators with non-zero determinant 
are always diagonlizable\footnote{Strictly speaking this is true for finite representations.}. This 
property is necessary to ensure the existence of a basis composed of eigen vectors for every 
observable. Because, properties of a system are by definition conserved under application of its 
symmetry group $G$, and observables are self-adjoint (hermitian) operators, representation of 
symmetry groups by state space must be unitary. This requirement enforces the necessity of definition 
of state space over $\mathbb{C}$. Although many groups are unitarily represented by vector spaces 
defined over real numbers, complete set of representations are realized on a complex field. 

Symmetry and similarity make sense only if there are at least two objects to be compared. Thus, 
a nontrivial Universe $U$ must have at least two non-empty nontrivial subsets $S$ (system), and 
$A$ (apparatus/observer). This fact explains why it seems that quantum mechanics needs the presence 
of an observer.

\subsection{Comparison with classical physics}
In Classical physics symmetries are passive, only reduce the number of degrees of freedom, and in 
many-body systems studied statistically, it makes distribution of some observables similar.
A classical system with $n$ observables has a configuration space isomorphic to $\bigotimes^n U(1) 
\cong \mathbb R^n$ irrespective of symmetries, except for conservation relations which impose 
constraintes. Consequently, at each moment the state of a system is a point in the 
configuration/phase space rather than a vector, and probabilities (observables) are $n-1$ simplexes 
in $\mathbb R^n$ and in general do not present a representation of symmetries of the system. 

\section{Inseparability of composite systems}
Properties of state space explained in section \ref {sec:state} along with axiom 
(\ref{postcomposite}), symmetries in particle physics, and universality of gravity, show that 
decomposition of system to components is never completely orthogonal and it is not possible to 
find two quantum subsystems $S_1$ and $S_2$ such that $S_1 \cap S_2 = \oslash$. This rule also applies 
to the Universe $U$ as a whole and means that it is an ensemble of intertwined vector spaces and is 
topologically open. Thus, for any system $S \subset U$ there is a system $S'$ such that 
$S \subset S' \subset U$. Consequently, locality and separability are approximations and Universe 
is inherently nonlocal, composite, and inseparable. 

As an application of this conclusion, we apply compositeness axiom to proposition of D.N. Page and 
W.K. Wootters~\cite{qmentangletime} to use conditional rather than absolute probabilities in 
quantum mechanics. Their aim is to provide a consistent description of evolution of quantum systems 
with diffeomorphism symmetry which have total Hamiltonian $\mathcal {H} = 0$. This proposition has 
been criticized~\cite{qmentantimecrit} mainly due to presence of an unobservable clock in the model. 
In the framework of symmetry description, the problem of this model can be recognized easily. 
One of their main assumptions is division of the Universe in two non-interacting parts which one 
is considered as system and the other as clock. According to postulate (\ref{postcomposite}) such 
a division is impossible, otherwise each part can be considered as a separate universe. Indeed 
modification of Page and Wootters proposal to solve the problem of unobservable 
clock~\cite{qmentantimemod} and its experimental realizations~\cite{qmentantimeex} include at 
least 3 subsystems which is consistent with minimum request of symmetry description. On the other 
hand, in the framework of symmetry description $\mathcal {H} = 0$ due to global diffeomorphism 
does not mean that Universe must be static, because it is by definition a composition of its 
subsystems which vary with respect to each others. One of these subsystems can be considered as 
a quantum clock and its entanglement with at least one other subsystem transfers the time to others. 
Such a clock is observable despite the absence of its direct interaction with all subsystems except 
one. Details of the operation of such clock will be discussed elsewhere.

\section{Measurement}
In standard description of quantum mechanics there are experiments in which it is not clear which 
quantity is measured. For example measurement of photon polarization with an interaction 
$\propto S_1 \times S_2$ (Dzyaloshinskii-Moriya interaction~\cite{dzmspinint,dzmspinint0}). 
Considering $S_1$ and $S_2$ as  of system and apparatus, respectively, the transverse plane 
to $\mathbf {S}_2$ can be considered as preferred plane for measurement. Then, if the direction of 
the spin to be measured is already known to be in the plane, for instance if system consist of 
photons propagating in vacuum parallel to $\mathbf {S}_2$, there is no preferred direction on the 
plane. In standard description of quantum mechanics this seems problematic because it is not clear 
which component of spin vector is measured. On the other hand, in symmetry description there is no 
ambiguity, and in all cases the measured observable is the unique generator of abelian subalgebra 
of $SU(2)$ which is independent of arbitrary definition of coordinates on the plane.

\subsection{Ambiguity of measurement basis}
Modern literature on foundation of quantum mechanics present measurement as entanglement between 
system and apparatus or environment~\cite{decohererev2,decohererev3}:
\be
\sum_\alpha s_\alpha |\alpha \rangle \otimes |a_0 \rangle \rightarrow 
\sum_\alpha s_\alpha |\alpha, a_\alpha \rangle \label{measentangle}
\ee
where $|a_0 \rangle$ is the initial state of the apparatus or environment and $|\alpha \rangle$ is 
an arbitrary basis for the system. In the framework of symmetry description, system and apparatus 
must have an interaction related to the symmetry represented by states $|\alpha \rangle$. Assuming 
for simplicity finite dimensional representations, dimension $M$ of the apparatus/environment initial 
representation must be larger than system dimension $N$. The entanglement by interaction breaks 
the $[N] \otimes [M]$ dimensional representation of system-apparatus to an $[N']$ dimensional 
representation. If $N' > N$ some outcomes would be equivalent. If $N' < N$ the apparatus will not 
be able to single out some states and they may create an interference. This reduction of state 
space, and thereby the symmetry, is usually explicit and related to the design of the apparatus, 
but there is no guarantee that $|\alpha, a_\alpha \rangle$ states be orthogonal to each others. 
In symmetry description this is not a problem because the right hand side of (\ref{measentangle}) 
presents the only physically meaningful entity, namely observable(s) associated to the symmetry 
in the representation realized by $|\alpha, a_\alpha \rangle$, even if these states are not eigen 
vectors.

\subsection{Decoherence and symmetry breaking} \label {sec:symmbrdeco}
Entanglement approach does not clarify the final breaking of degeneracies and decoherence. In 
symmetry description, the large number of degrees of freedom of environment present an infinite 
dimensional representation of symmetries, in general a degenerate representation containing many 
copies or equivalent subspaces. Entanglement reduces the state to:
\be
|\psi\rangle \otimes |E\rangle = |\psi, E\rangle = 
\sum\limits_{\alpha,~\epsilon^{\alpha}_i, i=0}^{i \rightarrow \infty} {f}(\alpha ,\{\epsilon^{\alpha}_i\}) | 
\alpha, \epsilon^{\alpha}_1, \epsilon^{\alpha}_2, \cdots, \epsilon^{\alpha}_i, \cdots\rangle 
\label{sysenvstate}
\ee
where $\alpha$ index characterizes a basis for symmetry representation of the system and 
$\epsilon_i$ presents equivalence class of representations of the system induced to 
environment/apparatus by interaction with system. Random selection of one member of these 
equivalence classes breaks the symmetry and its realization by the system becomes trivial. Thus, 
in contrast to classical systems immediate repetition of the measurement on the same system gives 
the same outcome. Because equivalence classes realize the same representation of symmetry as the 
system, they play the role of a grand canonical ensemble and one can define a probability for 
outcomes which would be determined by the symmetry/degeneracy of the state of system and completely 
independent of the environment. This shows the crucial role of symmetry and its breaking for 
measurement and decoherence.

\subsection{Symmetry breaking and phase transition}
The process of entanglement of system with environment is similar to a phase transition. The state of 
system-environment can be projected to an arbitrary basis similar to (\ref{sysenvstate}) and 
projection coefficients can be treated as a random field. The Sinai theorem~\cite{phtsinai} for 
phase transition in classical systems proves that for perturbations - due to noise or temperature - 
lower than a critical value, the probability distribution function of a random field to which a 
Hamiltonian is associated, is a limit Gibbs distribution for a single phase, and under suitable 
conditions the system has a probability close to 1 to be in one of the stable ground states. This 
theorem can be also applied to quantum systems, specially because the presence of equivalence 
classes of symmetry representations in the environment/apparatus is mathematically analogue to the 
classical case. Therefore if the setup of a measurement is suitable, a symmetry breaking does occur 
and it is a direct consequence of quantum entanglement presented in the r.h.s. of 
(\ref{sysenvstate}) which makes the system to approach a ground state and stays there. This theorem 
also predicts the existence of surfaces in coupling space in which multiple phases can coexist. 
For quantum systems they correspond to cases where interaction does not completely break the 
symmetry and produce an interference.

For instance, in a Stern-Gelach experiment, the environment is composed of photons of the magnetic 
field and spins of atoms in the detector. Although all these components contribute in 
system-environment state $|\psi, E\rangle$, only a limited number of photons get the chance to 
interact with the passing electron. The interaction Hamiltonian between two spins is proportional 
to $\hat{\vec{S}}_1.\hat{\vec{S}}_2$. Therefore, Hamiltonian of this ensemble depends on $s_is_j$ 
where $s$ is the eigen value of projection of spins. Considering the large number of photons, 
$\{s_i\}$ behave as a random field $s$ interacting with electron's spin. Although the state of the 
electron is initially a superposition of two polarizations, according to Sinai theorem, the photon 
spin field - the environment - pushes the electron spin - the system - to approach randomly to 
one of two minima of the Hamiltonian, corresponding to eigen value to the Cartan subalgebra of 
$\mathbb{Z}_2$ symmetry of polarized electrons along or opposite to detector's magnetic field, with a 
probability very close to 1 which breaks polarization superposition.

\section{Randomness, probability, and Born law} \label{sec:measure}
In symmetry description of quantum mechanics Born law for probabilities associated to possible 
outcomes of measurements can be proved rather than postulated. Following axiom (\ref{poststate}), 
state of a system contains all obtainable information about it. As probabilities $f_i$ are by 
definition positive or zero real values, they must depend on absolute value of projection 
coefficients of the state on the eigen vectors of the observable operator $|a_i|^2$ with 
$\sum_i^n f_i (|a_i|^2) = 1$. In addition, without loss of generality and due to projective nature 
of state space, states can be normalized such that $\sum_i^n |a_i|^2 = 1$. Considering the state 
space of a system composed of two similar part:
\be
|\psi\rangle = \frac{1}{2}(|\psi_1 \rangle \otimes |\psi_2\rangle + |\psi_2 \rangle \otimes 
|\psi_1\rangle) = \sum_i a_i^2 |i\rangle \otimes |i\rangle + \sum_{i \neq j} a_i a_j 
(|i\rangle \otimes |j\rangle + |j\rangle \otimes |i\rangle) \label{compsys}
\ee
Symmetrization over two subsystems means that due to similar preparation they are indistinguishable. 
From this expression we conclude that the probability for subsystems to be in $(i,j)$ state is 
$f(|a_ia_j|^2)$. On the other hand, because these subsystems are prepared independently, the mutual 
probability for the system to be in $(i,j)$ state is $f(|a_i|^2)f(|a_j|^2)$. Therefore:
\be 
f(|a_i|^2)f(|a_j|^2) = f(|a_i|^2|a_j|^2) \label{coeffprop}
\ee
It has been proved~\cite{hardymodel} that the only function with such property is a positive 
power-law. Finally, by applying (\ref{coeffprop}) to summation properties, one finds the Born law:
\be
f(|a_i|^2) = |a_i|^2
\ee
This relation also proves that state space must be a Hilbert space and show that the reformulation 
of quantum axioms based on symmetry is equivalent to standard description. Nonetheless, it gives 
a more logical structure to the model and its predictions

\section{Outline}
Introduction of symmetry in axioms and fundamental structure of quantum mechanics is not just 
for clarification of some subtleties and confusing predictions. Symmetry is a fundamental concept 
in construction of physical Universe and its inclusion in foundation of quantum mechanics is most 
probably crucial and necessary for building an abstract model of the Universe and its content. An 
evidence for this is the central role of gauge symmetry in particle physics and diffeomorphism 
symmetry of spacetime. We highlighted a couple of consequences of symmetries for global view of 
cosmology in this proceedings, but many others must be explored and understood, notably in the 
path to discovery of a unified model of quantum process which includes gravity.

\end{document}